\documentclass[conference]{IEEEtran}
\IEEEoverridecommandlockouts
\usepackage{cite}
\usepackage{amsmath,amssymb,amsfonts}
\usepackage{algorithmic}
\usepackage{graphicx}
\usepackage{textcomp}

\usepackage{xcolor}
\usepackage{tcolorbox}
\usepackage{tabularx}
\usepackage{color, colortbl}
\usepackage{soul}
\usepackage{booktabs}
\usepackage{comment}
\usepackage{makecell}
\usepackage{hyperref}
\usepackage{amsthm}
\usepackage{tikz}
\usepackage{listings}
\usepackage{varwidth}
\usepackage{xurl}
\usepackage{balance}
\usepackage{dblfloatfix} 
\usepackage[moderate,tracking=normal]{savetrees}
\usepackage{svg}

\definecolor{myred}{RGB}{174, 0, 0} 
\definecolor{myellow}{RGB}{252, 196, 84}

\newcommand{\toolname}[1]{\emph{KubeFence}}

\definecolor{DarkGreen}{RGB}{65, 148, 87}
\lstdefinelanguage{yaml}{
    morekeywords={apiVersion, kind, metadata, spec, initContainers, containers, volumeMounts, volumes, name, image, command, mountPath, subPath, emptyDir, type, data, labels, dict, app, chart, release, heritage, PGUSER, PGPASSWORD, tracking, enabled, replicaCount, host, pullSecrets, repository, registry, postgreSQL, arch, containerSecurityContext, runAsNonRoot, template, securityContext, hostPath, path, imagePullPolicy, ports, container},
    sensitive=true,
    morecomment=[l]{\#},
    commentstyle=\color{DarkGreen}\bfseries,
    morestring=[b]",
    morestring=[b]',
    stringstyle=\color{black},
    showstringspaces=false,
    basicstyle=\ttfamily\scriptsize,
    keywordstyle=\color{purple}\bfseries
}

\tcbuselibrary{skins}
\NewTotalTColorBox[auto counter]{\Definition}{ +m }{ 
    notitle,
    colback=blue!3!white,
    frame hidden,
    boxrule=0pt,
    enhanced,
    sharp corners,
    borderline west={4pt}{0pt}{blue!20!white},
     boxsep=2pt,  
    left=7pt,    
    right=2pt,   
    top=2pt,     
    bottom=2pt,  
}{
    \textcolor{green!50!black}{
        \sffamily
    }%
    #1
}

\begin{document}

\title{ \toolname{}: Security Hardening of the \\ Kubernetes Attack Surface}

\author{
    \IEEEauthorblockN{Carmine Cesarano, Roberto Natella}
    \IEEEauthorblockA{Universit\`a degli Studi di Napoli Federico II, Italy\\ 
\{carmine.cesarano2, roberto.natella\}@unina.it}
}

\maketitle
\begin{abstract}
Kubernetes (K8s) is widely used to orchestrate containerized applications, including critical services in domains such as finance, healthcare, and government. However, its extensive and feature-rich API interface exposes a broad attack surface, making K8s vulnerable to exploits of software vulnerabilities and misconfigurations. Even if K8s adopts role-based access control (RBAC) to manage access to K8s APIs, this approach lacks the granularity needed to protect specification attributes within API requests. 
This paper proposes a novel solution, \toolname{}, which implements finer-grain API filtering tailored to specific client workloads. \toolname{} analyzes Kubernetes Operators from trusted repositories and leverages their configuration files to restrict unnecessary features of the K8s API, to mitigate misconfigurations and vulnerabilities exploitable through the K8s API. The experimental results show that \toolname{} can significantly reduce the attack surface and prevent attacks compared to RBAC.
\end{abstract}

\begin{IEEEkeywords}
K8s, Attack Surface, API Filtering
\end{IEEEkeywords}

\section{Introduction}
Kubernetes (K8s) \cite{kubernetes} has become the mainstream platform for orchestrating containerized applications, enabling scalable deployment and management across distributed environments. Its adoption spans various critical domains, including finance, healthcare, and government services \cite{CNCF_survey, k8s_usecases}, where security is crucial. As K8s becomes integral to these sensitive areas, ensuring the security of its components, particularly the K8s API, is of the highest importance.  

K8s is a project with a huge codebase and a large, complex interface toward clients \cite{openhub_kubernetes}. This interface provides convenience of use and feature-richness, but it also represents an ``attack surface'' that exposes the system to security attacks \cite{theisen2018attack}. If these features are provided by vulnerable code, they can be exploited by malicious users to pursue attacks. In the case of K8s, attackers leverage vulnerabilities to run unauthorized workloads, such as cryptojacking and botnets \cite{crypto_incident}; moreover, attackers can violate the isolation between tenants in the infrastructures, such as disrupting applications and stealing or damaging their data \cite{yang2023take}.

To avoid such security risks, regulatory frameworks and security agencies are recommending the adoption of secure design practices  
\cite{cisa2023securitybydesign,ncsc2020securitybydesign,ec2024cra}. In particular, software systems should adopt the ``principle of least privilege'', by minimizing the attack surface to only provide access to the strict minimum of features and resources \cite{saltzer1975protection,smith2012contemporary}. In the case that a user behaves maliciously, they should be prevented from accessing unnecessary features that could be exploitable \cite{yang2023take}. 

In practice, this approach is quite challenging to apply in complex systems such as K8s. K8s provides a REST API to manage resources, such as \textit{Pods}, \textit{Services}, \emph{Deployments}, and several others. 
It adopts \emph{role-based access control} (RBAC) \cite{kubernetes_rbac} to manage API calls that access these resources. However, K8s builds more complex abstractions on top of the REST API. When a resource is configured through the REST API, the API takes in input complex data structures (as YAML payload of a request) to describe the ``desired state'' of the resource (\emph{specification}). These data structures can contain attributes to use advanced features, such as access to resources on the host machine and other special permissions. These features represent a risky attack surface that can be exploited in the case of software vulnerabilities in K8s.

We argue that RBAC does not suffice to secure K8s, since it only provides access control on resources as a whole, but lacks control over specific features in the specification of these resources. 
For example, while RBAC can allow users to manage \textit{Pod} resources and disable access to \textit{Deployment} resources, it cannot restrict the values of fields within the specification of a \textit{Pod}. 
These features can be triggered by a malicious client in order to exploit vulnerabilities in K8s.

In this paper, we analyze the extent of the K8s attack surface and how to harden it. We initially present an analysis of CVE vulnerability records \cite{k8s_cve_feed} for the K8s project, in which we found that CVEs are only exploitable through specific features of the K8s API interface. Then, we present a solution (\toolname{}) to provide finer-grain control of the K8s attack surface to harden K8s against exploits and misconfigurations. Our solution analyzes applications that use K8s from trusted repositories (\emph{Kubernetes Operators \cite{k8s_operator}}), and generates security policies that restrict API access to only the resources explicitly required by the application.

In our evaluation, we defined a catalog of malicious K8s specifications for testing \toolname{}, based on known CVEs and common misconfigurations. We applied \toolname{} on several popular applications. Our solution was able to mitigate all tested CVE exploits and misconfigurations. Moreover, we found that \toolname{} achieved an average of $35$\% reduction in the attack surface compared to RBAC, while introducing an acceptable performance impact for non-realtime workloads. In particular, we measured an average overhead of $50$ ms for cluster management operations ($\sim$21\%). Since \toolname{} only applies to cluster management, once containers have been deployed, it does not affect container execution.

The main contributions of this work are: 
\begin{itemize}
    \item We analyze K8s CVEs and found that vulnerable code is only exercised when specific fields are used in API requests.
    \item We design \toolname{} to generate and enforce security policies for the K8s API interface. \footnote{Available as open-source at: \url{https://github.com/dessertlab/kubefence/}.}
    \item We present a catalog of attacks against the K8s API interface, based on CVEs and on common misconfigurations.
    \item We evaluate the effectiveness of \toolname{} at reducing the attack surface, at mitigating CVEs and misconfigurations, and at achieving efficiency with minimal runtime overhead.
\end{itemize}

\begin{figure}[t]
  \includegraphics[width=\columnwidth]{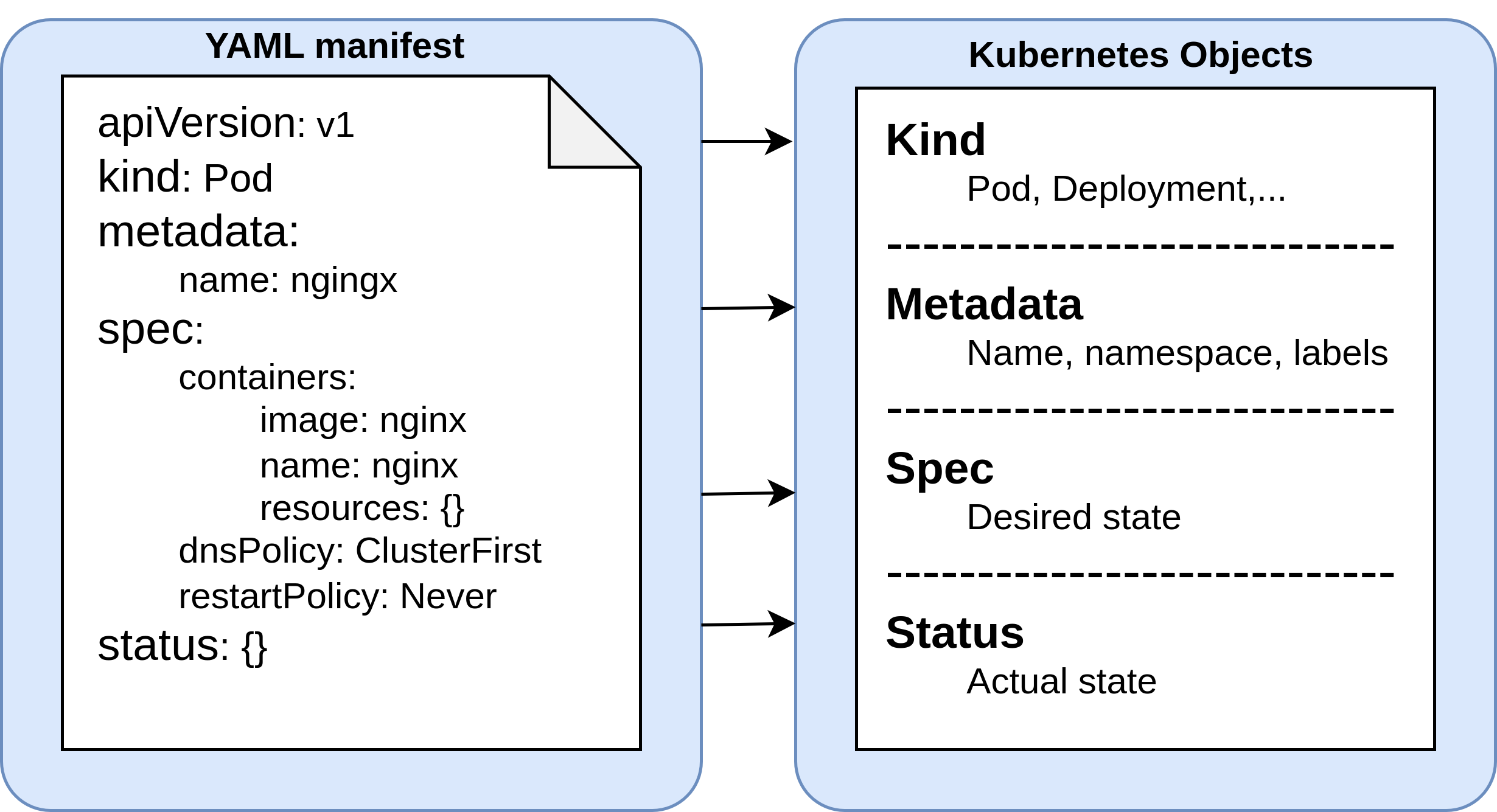}
  \caption{An example of YAML Manifest to configure a Kubernetes Object.}
  \label{fig:manifest_yaml}
\end{figure}

\section{Background}
\subsection{Kubernetes}
\label{sec:k8s_background}
Kubernetes (K8s) is an open-source platform for automating the deployment, scaling, and management of containerized applications. At its core, Kubernetes abstracts underlying physical resources into logical entities called \textit{Kubernetes resources}, such as Pods (for workloads), Services (networking), Volumes (storage), ConfigMaps (configuration), and Secrets (security). These resources are managed within a cluster, consisting of worker nodes orchestrated by a control plane. The K8s API Server is the central management component, responsible for handling operations within the cluster. It exposes a RESTful API that allows users to interact with the cluster by sending HTTP requests to create, modify, and delete resources. The API Server supports HTTP verbs like \textit{get}, \textit{post}, \textit{put}, and \textit{delete} to manipulate Kubernetes resources. Each resource is represented as a \textit{Kubernetes Object}, typically defined using a declarative manifest file in YAML or JSON format. The manifest specifies the desired state of a resource, which the K8s control plane works to reconcile with the current state. Kubernetes Objects generally contain two primary nested properties, which are \texttt{spec} and \texttt{status}, to describe the desired and current state, respectively. This structure is typical for objects like Pods, Deployments, and Volumes, as shown in Figure \ref{fig:manifest_yaml}. An example of a desired state is to bring up a given number of container replicas. Objects like Secrets and ConfigMaps, omit these fields, focusing on storing sensitive data or configuration in the \texttt{data} field. The K8s API exposes a set of endpoints for each resource. When a user defines a resource in a manifest, specifies all configurable fields, and applies this configuration to the cluster, an HTTP request is triggered to the relevant API endpoint, including all the resource configurations in the payload. This workflow allows users to configure resources declaratively by specifying their intent, while the API server translates these configurations into actual cluster state changes. However, these exposed API endpoints and configurable fields contribute to a significant attack surface, as discussed in Section \ref{sec:motivation}.

\subsection{Kubernetes RBAC}
To mitigate security risks, K8s provides an RBAC access control mechanism \cite{kubernetes_rbac}. RBAC defines which users, groups, and service accounts can perform specific actions on resources, such as viewing, creating, modifying, and deleting them. RBAC policies are defined using YAML manifests through four kinds of Kubernetes Objects: \textit{Role}, \textit{ClusterRole}, \textit{RoleBinding}, and \textit{ClusterRoleBinding}. A Role or ClusterRole object contains rules that specify a set of permissions, while RoleBinding and ClusterRoleBinding objects grant the permissions defined in a role to a user or set of users. This is particularly relevant in multi-user K8s clusters, where developers should be restricted to working with designated objects, preventing them from accessing others.

While RBAC provides a structured approach to access control, it has limitations in terms of granularity.  RBAC policies do not inspect the contents of K8s resource specifications in input to the API. This means that even if the user is restricted to only access specific K8s resources, they can still abuse or exploit all of the features available for that resource. Therefore, a more granular enforcement mechanism capable of inspecting and filtering API requests at a deeper level to block potential misconfigurations and exploits, as discussed in Section~\ref{sec:motivation}. 

\begin{figure}[t]
  \includegraphics[width=\columnwidth]{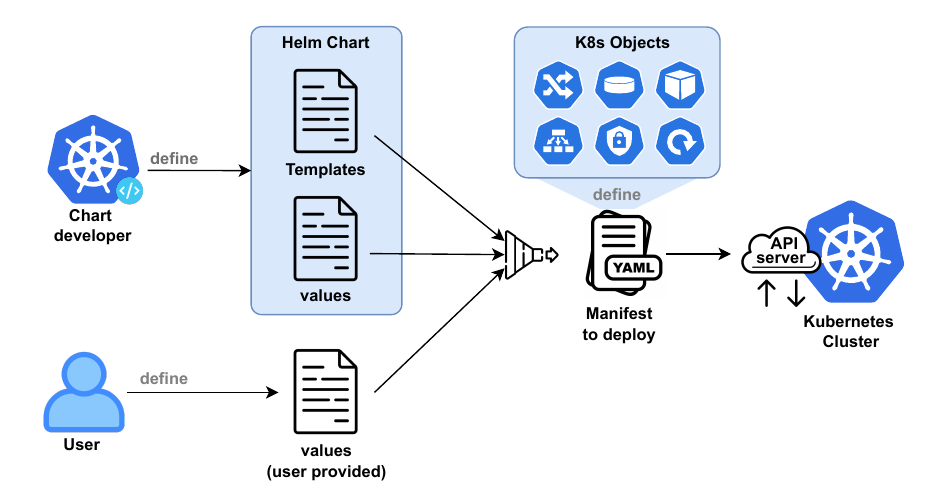}
  \caption{Helm Template Processing.}
  \label{fig:helm_template}
\end{figure}

\subsection{K8s Operators and Helm templates}
\label{sec:K8s_operators}
Kubernetes relies on \emph{controllers} to continuously monitor and reconcile the desired and current states of cluster resources. Built-in controllers are preconfigured to handle standard Kubernetes resources and provide basic features, such as autoscaling and self-healing. However, for more complex operations that extend beyond the capabilities of built-in controllers, users must adopt custom controllers. These ad-hoc controllers, known as \textit{Kubernetes Operators \cite{k8s_operator}}, are K8s API clients that automate advanced lifecycle management tasks for stateful and specialized applications. Operators continuously monitor and adjust the application state in a control loop. For instance, if the user specifies that an application should maintain three replicas, the Operator constantly checks the cluster state. If it detects that one replica has failed, it automatically triggers a new deployment to restore the desired count. This capability allows operators to handle both Day-1 (installation, configuration) and Day-2 (updates, scaling, monitoring) operations, reducing manual intervention.

Operators can be implemented in various ways, using the Go language, Ansible, and Helm \cite{redhat_operators}. Among these, Helm-based operators are by far the most common. The widespread adoption of Helm is evident in catalogs such as Artifact Hub \cite{artifact_hub} and OperatorHub \cite{operatorhub}, which lists hundreds of Helm-based Operators already distributed for production use, spanning applications such as databases, monitoring tools, and CI/CD pipelines.

Helm is the de facto package management solution for K8s \cite{zerouali2023helm}, to simplify the process of handling complex K8s resources to package, configure, and deploy applications. Helm packages, known as \textit{charts}, provide \texttt{templates} files, that are created by chart developers. These templates provide definitions of K8s resources to run an application. 
These templates consist of fixed parts and of placeholders for configurable values, as shown in Figure \ref{fig:helm_template}. Defaults for the configurable values are typically included in a separate \texttt{values} file, in order to provide an initial configuration for the K8s resources. Users of the K8s operator can customize and override to meet specific workload requirements. This flexibility allows users to deploy the same application across different environments with minimal changes \cite{henning2021reproducible}. 

\begin{figure}[b]
\begin{lstlisting}[language=yaml]
apiVersion: v1
kind: Secret
metadata:
  name: {{ template "mlflow.fullname" . }}-env-secret
  labels:
    dict: Dict
    app: {{ template "mlflow.name" . }}
    chart: {{ template "mlflow.chart" . }}
    release: {{ .Release.Name }}
    heritage: {{ .Release.Service }}
type: Opaque
data:
  dict: Dict
{{- if .Values.backendStore.postgres.Enabled }}
  PGUSER: {{ .Values.backendStore.postgres.user }}
  PGPASSWORD: {{ .Values.backendStore.postgres.password }}
{{- end }}
\end{lstlisting}
\caption{Helm \textit{Template} for a \textit{Secret} resource.}
\label{fig:helm_template_secret}
\end{figure}

When deploying an application, Helm processes the template by combining them with values to generate complete K8s manifests. Beyond simple value assignment, Helm templates support advanced conditional logic through directives such as \texttt{if-else} or \texttt{range}. These directives enhance template flexibility, enabling developers to include or exclude fields, iterate over collections, or conditionally populate fields based on the provided values (e.g., enabling optional configurations, as shown in Figure~\ref{fig:helm_template_secret}). These manifests are then submitted to the K8s API Server to create or update resources. In practice, Helm templates constrain the inputs that are sent to the K8s API Server, since the user does not change the fixed parts of the templates. We leverage this insight to harden the attack surface of the K8s API Server, as discussed in Sec. \ref{sec:kubefence_design}.

\section{Motivation}
\label{sec:motivation}
The flexibility and extensibility of K8s, while providing significant advantages for deployment and scaling, also introduce substantial risks. The attack surface exposed by the K8s APIs is particularly concerning, as \emph{malicious API requests} can abuse features and exploit vulnerabilities of K8s. 
This section discusses these security threats and motivates the need for finer-grain security controls to mitigate them.

\subsection{Misconfigurations of a K8s cluster}
K8s is not inherently secure by default. Proper configuration is crucial to maintaining a secure environment, but the complexity of this task often leads administrators to prioritize ease of deployment over rigorous security practices. This can result in security misconfigurations that inadvertently weaken the security of the cluster \cite{nsacisa2022hardening}. For example, running containers with elevated privileges or misapplying resource limits can expose critical resources and escalate privileges. Empirical studies \cite{misconfiguration2023} have shown that common configuration errors, such as overly permissive network policies or default access settings, can leave clusters vulnerable to exploitation.

Malicious users can leverage these misconfigurations to abuse the cluster. For example, if the cluster runs containers with high privileges, and the user omits the \texttt{runAsNonRoot} specification attribute, the user can escalate privileges. Another example is to leave service accounts enabled (e.g., \texttt{defaultServiceAccount}), which provides the user with permission to access the K8s API in every namespace. Finally, some functionalities require careful configuration to avoid introducing weaknesses. For example, inadequate TLS/SSL settings can expose communication channels to interception.

These issues often do not stem from flaws in K8s itself, but rather from how the system is configured by system administrators. When misconfigured, a K8s cluster can quickly become an attractive target for attackers. 
It is also important to note that RBAC does not provide control over potential abuses of such features.

\begin{figure}[t]
\begin{lstlisting}[language=yaml]
apiVersion: v1
kind: Pod
spec:
  initContainers:
  - name: busybox
    image: "busybox"
    command: ["ln", "-s", "/", "/mnt/data/symlink-door"]
    volumeMounts:
    - name: test-vol
      mountPath: /test
  containers:
  - name: my-container
    image: "ngingx"
    volumeMounts:
      - mountPath: /test
        name: my-volume
        subPath: symlink-door
  volumes:
  - name: my-volume
    emptyDir: {}
\end{lstlisting}
  \caption{Malicious K8s API request triggering CVE-2017-1002101.}
  \label{fig:subpath_misconfiguration}
\end{figure}

\begin{figure*}[t]
  \includegraphics[width=\textwidth]{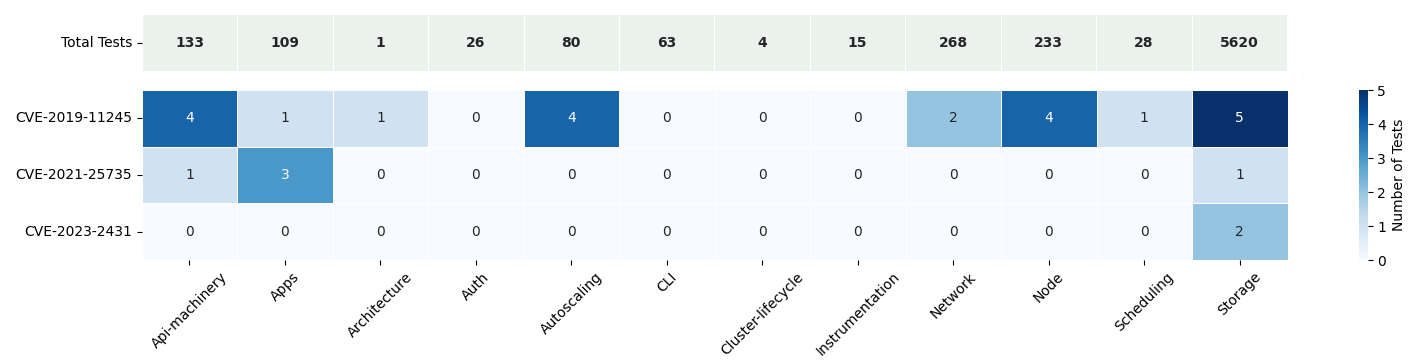}
  \caption{Number of e2e tests in each category that interact with vulnerable files associated with a CVE.}
  \label{fig:CVE_heatmap}
\end{figure*}

\subsection{Software Vulnerabilities in the K8s codebase}
\label{sec:software_vulnerabilities}
Beyond misconfigurations, K8s itself is susceptible to vulnerabilities in its codebase. Numerous CVEs have been found in the recent past, and more are likely to occur in the future due to the complexity of the K8s project. Some of these CVEs can be directly exploited through the K8s API, by injecting malicious input values in object specifications. These exploits can cause disruption of cluster operations, privilege escalation, and unauthorized access to sensitive data. 

For example, in K8s clusters prior to version 1.9.4, the vulnerable \texttt{subPath} feature can be exploited by attackers to access sensitive directories on the host filesystem (\href{https://nvd.nist.gov/vuln/detail/cve-2017-1002101}{CVE-2017-1002101}). As illustrated in Figure \ref{fig:subpath_misconfiguration}, this vulnerability can be exploited by sending a maliciously crafted API request to deploy a K8s Pod. Specifically, an init container creates a symbolic link to the root directory of the host, and the main container then mounts this symlink as a \texttt{subPath}, granting access to the host filesystem. 

These vulnerabilities demonstrate that restricting user permissions at a high level, as does RBAC, is insufficient. Limiting access to certain resources (e.g., by denying access to other resources except pods) does not prevent attackers from manipulating specific configuration parameters of unrestricted resources to exploit underlying vulnerabilities. 
The limitations of RBAC are intrinsically due to its conceptual model, which is coarsely defined around ``roles'' and ``resources'', since it is designed for manual definition and review by administrators. Even if a finer-grained RBAC existed, it would introduce excessive complexity for this use case. 
Thus, an automated and more detailed filtering of parameters within API requests is necessary to reduce the attack surface.

\subsection{Attack Surface across Workloads}
\label{sec:CVE_mapping}
We have seen that the K8s API is exposed to attacks against K8 misconfigurations and vulnerabilities. We hypothesize that, in practice, many of these vulnerabilities and misconfigurations can be triggered by exploiting specific features of the K8s API that are not required by many users. If this hypothesis holds, it would be possible to prevent attacks by blocking unnecessary features in a workload-specific manner, thus reducing the attack surface of K8s.

To test this hypothesis, we analyzed past vulnerabilities in K8s, and which features can trigger them from the API. 
We first analyzed the official K8s Vulnerability Database \cite{k8s_cve_feed}, covering all entries from July 2016 to December 2023. This effort yielded a total of 49 CVEs with CVSS scores ranging from 2.6 (low severity) to 9.8 (high criticality). By examining the source code files modified by the patches for these CVEs, we were able to map each vulnerability to the corresponding affected K8s components. These components span a wide range of functionalities, including admissions controllers, kubelet, API server, etcd, kubectl, scheduler, networking, storage, the legacy cloud provides support and security features. We provide the full mapping in the project repository.

Then, we adopted a set of \emph{workloads} to exercise the features exposed by the K8s API. 
We chose K8s end-to-end (e2e) tests for this purpose \cite{e2e_tests}. e2e tests were selected because they perform realistic interactions with the K8s API, by deploying resources and managing configurations, as seen in production environments. Unlike simple resource operations (e.g., create, delete), these workloads involve complex API requests that selectively trigger different K8s features. For example, an e2e test that manages CustomResourceDefinitions (CRDs) \cite{e2e_test_example} uses service names, ports, and conversion strategies, which exercise specific parts of the K8s codebase. This makes e2e tests ideal as workloads to analyze the relationship between vulnerabilities and features of the K8s API. If a vulnerability can only be triggered through a specific feature, we expect to see that only a small minority of tests cover the vulnerable code.

We selected all available e2e tests across $12$ different categories (e.g., networking, storage, scheduling, autoscaling, etc.), excluding tests designed for \textit{Windows} environments because our testbed is built on Linux, and tests in the \textit{disruptive} category, as their focus is on resilience and fault tolerance rather than functional testing. In total, we selected $6,580$ e2e tests. It is important to note that the distribution of e2e tests is not uniform across K8s components. This depends on the richness of configuration parameters available for some components (e.g., storage), where the higher number of tests reflects a greater variety of associated workloads. Therefore, we chose not to sample the tests, and to include the full test suites. We need to consider this imbalance when interpreting the results. Before execution, we instrumented the codebase to collect code coverage data, allowing us to track which lines of source code were accessed by each test \cite{kubernetes_coverage}. We cross-reference these data with the vulnerable files identified earlier. 

Figure \ref{fig:CVE_heatmap} illustrates the total number of e2e tests grouped by category (columns of the matrix), where each test category interacts with different parts of the K8s API and requires different K8s resources. In addition, the figure shows, as a heatmap, the number of tests that cover vulnerable code linked to 3 CVEs (rows of the matrix). 
We found that vulnerable code is covered only by a very small minority of workloads. 
For example, in the case of CVE-2023-2431, only two workloads from the storage e2e tests cover the vulnerable code. The vulnerable code for the other 46 CVEs is not covered by any of the tests, and not shown in the figure for the sake of brevity. In total, only $29$ out of $6,580$ tests (i.e., less than $0.5\%$) exercised a vulnerable part of the codebase. 
Even if the distribution of e2e test across categories is skewed towards storage, which accounts for the majority of tests, the skew does not undermine this conclusion. We find that, even when excluding the largest category, vulnerable code is covered by only $21$ out of $960$ tests (i.e., around $2\%$).

In conclusion, we investigate the overlap between the features commonly used by workloads and the ones exploitable by attacks. Since the overlap is small in practice, disabling unnecessary features when executing a particular workload can thwart many attacks. This preliminary analysis shows that by enforcing API access controls tailored to specific workloads, we can significantly limit exposure to (potentially vulnerable) components that are not necessary, thereby reducing the risk of exploitation. This workload-specific filtering approach can thus minimize the K8s attack surface, addressing a critical gap in the existing coarse-grained RBAC model.

\subsection{Threat Model}
Our threat model assumes attackers who have gained control over the cluster and can execute commands against the K8s API. These attackers include compromised users with stolen credentials, over-privileged users, and other types of insider threats \cite{threat_model}. Such attackers may attempt to escalate privileges to gain full control over K8s resources (e.g., Pods, Deployments, Services), access the underlying hosts in the cluster, and disrupt provided services. To achieve these objectives, they can misuse the API to deploy malicious resources or reconfigure existing ones with harmful \textit{specifications}, in order to leverage cluster misconfigurations or exploit vulnerabilities in the K8s codebase. An example of attack based on this threat model was described in Section \ref{sec:software_vulnerabilities}.

Other security threats, such as physical attacks on infrastructure (e.g., compromising the physical machine hosting the etcd database) and supply chain attacks (e.g., targeting container images or CI/CD pipelines) are considered out of scope for this work. In addition, we do not consider volumetric denial-of-service attacks, such as API flooding with high-volume request patterns. Our threat model still considers ``non-volumetric'' DoS that may be caused by malicious API requests from CVE exploits, which can disrupt workloads and cause service unavailability.

\Definition{
To sum up, we discussed how the existing K8s security model based on RBAC is insufficient to prevent abuses of cluster misconfigurations and exploitation of vulnerabilities. The analysis of K8s vulnerabilities showed that they can be exploited only through specific features of the K8s API. Therefore, we design \toolname{} for fine-grain filtering of K8s API requests to reduce the K8s attack surface.
}


\section{Challenges}
Our \toolname{} solution is based on the fundamental idea of leveraging K8s Operators to obtain strict security policies. 
K8s operators are becoming more and more popular, and represent a paradigm shift to manage clusters. With operators, developers implicitly encode choices on which features they use. However, the current security model of K8s does not leverage this as an opportunity to restrict the large attack surface. There are technical challenges in doing so.

\toolname{} generates security policies from Operator configurations, commonly defined using Helm Charts. These charts bring flexibility through templating, which includes conditionals, loops, data types, and user-defined overrides (Sec. \ref{sec:k8s_background}). 
The challenge lies in ensuring that security policies generalize across all valid configurations that can be derived from charts. To address this, \toolname{} systematically explores the configuration space of an Operator, by identifying valid variants of the configuration, while restricting them to specific attributes and values where possible.

In addition, K8s API requests contain deeply nested objects with flexible, optional fields, making precise validation challenging. Traditional RBAC only checks K8s resources and actions, whereas fine-grained enforcement requires inspecting the full request structure. A flat-object approach would overlook dependencies between nested fields, enabling attackers to bypass restrictions. To address this, \toolname{} employs a tree-based validation mechanism that mirrors the hierarchical structure of Kubernetes API requests.

Finally, the architecture of \toolname{} should fit between clients and the K8s cluster, by ensuring that API requests cannot bypass security validation, with minimal resource and performance overheads.

\begin{figure}[t]
  \includegraphics[width=\columnwidth]{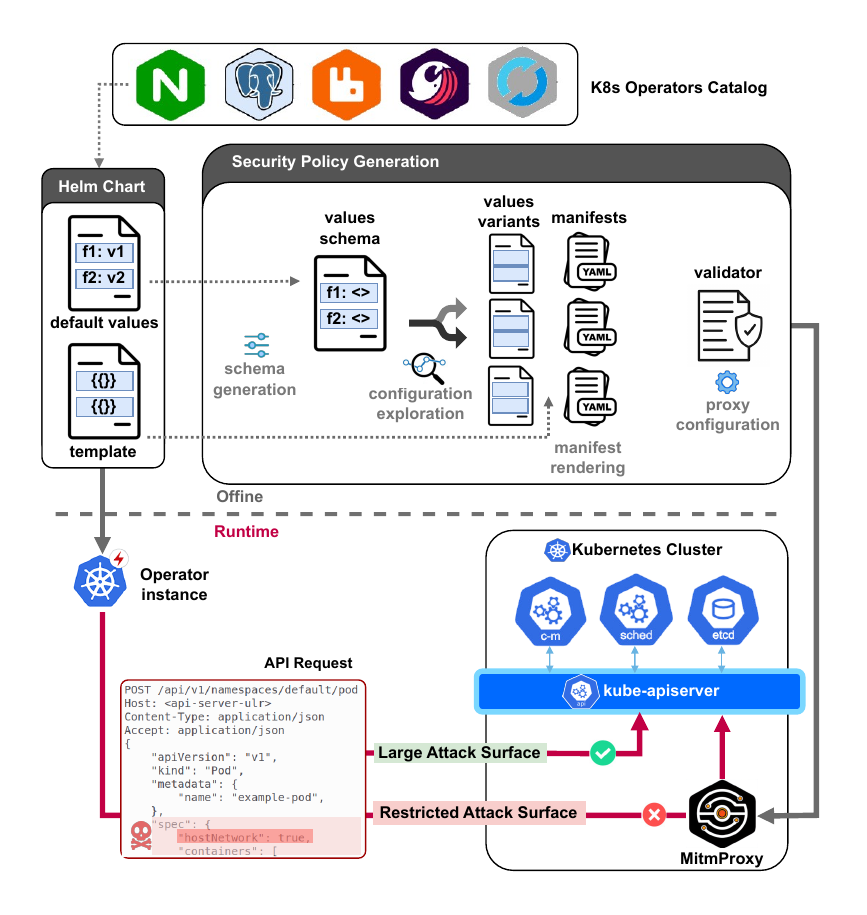}
  \caption{KubeFence overview.}
  \label{fig:kubefence_overview}
\end{figure}

\section{KubeFence Design}
\label{sec:kubefence_design}
\toolname{} is a proxy-based enforcement mechanism designed to automatically generate and enforce fine-grained API security policies, tailored to specific K8s workloads. A \textit{security policy}, in our context, defines allowable resource specifications, which restricts the attack surface by filtering API requests that include unnecessary attributes. These policies are represented as \textit{validators}, a machine-readable format used by our proxy to check API requests. The proposed approach is tailored for Helm-based K8s operators, which are widely used to manage complex Kubernetes configurations and require careful security analysis \cite{zerouali2023helm}.

Unlike static security policy checkers that only define allowed resource specifications, \toolname{} enforces these policies dynamically at runtime. By intercepting and validating every API request during cluster operations, \toolname{} ensures that only authorized configurations are applied, effectively preventing unauthorized API interactions that could bypass static security measures.

\toolname{} seamlessly integrates into the existing Kubernetes ecosystem with minimal disruption to the workflow of administrators and developers. Its operation involves three primary steps, as depicted in Figure \ref{fig:kubefence_overview}: (1) analyzing K8s workloads and their Helm charts to identify required K8s objects and enumerate their potential configurations; (2) generating security policies based on this analysis and configuring an API proxy to enforce them; (3) intercepting incoming API requests, validating them against the defined policies, and blocking any that deviate from the allowed configurations. By automating policy generation and enforcement, \toolname{} reduces the manual effort required to secure Kubernetes deployments while enhancing fine-grained protection against insider threats.

\subsection{Generation of Security Policies}
Writing security policies is a complex and error-prone process, especially for complex systems such as K8s. Inaccurate or incomplete policy definitions can leave clusters vulnerable to overly permissive access control.

To address this challenge, \toolname{} automates the generation of fine-grained security policies by analyzing Helm charts (default values and templates) as input. 
The goal of \toolname{} is to produce a consolidated policy in the form of a \textit{validator}, that is, a reference schema for the validation of incoming API requests. \toolname{} ensures that the K8s object specification in an API request, i.e., a manifest (see also \figurename{}~\ref{fig:helm_template}), complies with the fields and values of the schema. 
The schema defines all allowable configurations for each K8s resource defined by the Helm chart. 
An API request that uses an attribute not included in the schema can be blocked, since it is unnecessary according to the Helm chart. Similarly, if the schema assigns a specification attribute with a fixed value, or a value taken from a small set, API requests that use any value outside this defined range can be blocked.

However, there are several aspects that need to be handled for accurate security policies. (1) Conditional logics in Helm charts dynamically vary the structure of the specification based on user-defined values (as in Figure~\ref{fig:helm_template_secret}). This variability means that many potential configurations generated by these conditions need to be accounted for in the policy. (2) Moreover, while Helm charts often provide default values that fix the structure and content of manifests, users can still override these defaults with custom values (as illustrated in Figure~\ref{fig:helm_template}). Thus, policies should not rely solely on the templates of the Helm charts but should account for such overrides from the user of the K8s operator. (3) Finally, the K8s specification includes critical attributes that are recommended by security best practices (such as \texttt{runAsNonRoot}) but that may be omitted in the Helm charts. Policies must ensure these critical attributes are enforced in API requests, regardless of their presence in the Helm charts.

These aspects make policy generation more nuanced. 
\toolname{} manages them by exploring the configuration space represented in the Helm charts, to ensure that policies cover all legitimate API requests, while guarding against potential API misuses. 
The policy generation process is divided into four phases, detailed below.

\vspace{2mm}
\noindent
\textbf{\textit{Generation of values schema}}.
This phase analyzes values of K8s resource specifications in the Helm charts, in order to identify the domain of every field. The value schema will serve as a basis for exploring the configuration space in the next phase. 

\toolname{} performs a transformation of the default values to produce a structured \textit{values schema}. 
This transformation aims to: (1) Replace static values with placeholders representing data types or valid ranges, such as \texttt{bool}, \texttt{string}, \texttt{int}, \texttt{IP}, \texttt{[list]}, and \texttt{\{dict\}}, using regex-based substitution. 
(2) Replace enumerative fields replaced with lists including all valid options, extracted from annotations in the values file. (3) Lock predefined safe constants to fields critical to security, according to best practices for K8s resource specifications. For example, \texttt{securityContext.runAsNonRoot} can be locked to \textit{true} \cite{sec_k8s_guidelines}. Similarly, fields like \textit{registry} and \textit{image name} can be restricted to trusted values to mitigate risks like typosquatting attacks \cite{liu2022exploring}. Thus, security-sensitive fields are locked with safe constants rather than placeholders, and any missing critical field is explicitly added. Figure~\ref{fig:generated_values} demonstrates an example of this process applied to the MLflow Operator.

\begin{figure}[t]
\begin{lstlisting}[language=yaml]
# Default Values File         |# Values Schema
image:                        |image:
 registry: docker.io          |  registry: docker.io
 repository: bitnami/mlflow   |  repository: bitnami/mlflow
 pullSecrets:                 |  pullSecrets: [list]
    - name: secret-1          |
    - name: secret-2          |
tracking:                     |tracking
 enabled: true                |  enabled: bool
 replicaCount: 1              |  replicaCount: int
 host: "0.0.0.0"              |  host: IP
 containerSecurityContext:    |  containerSecurityContext:
   runAsNonRoot: true         |    runAsNonRoot: true
# postgresql.arch             |
# standalone` or `repl        | 
postgreSQL:                   |postgreSQL:
 arch: standalone             |  arch: standalone, repl
\end{lstlisting}
\caption{Schema generation from the \textit{default Values} file used in the MLflow.}
\label{fig:generated_values}
\end{figure}

\vspace{2mm}
\noindent
\textbf{\textit{Exploration of the configuration space}}.
The values schema produced in the previous phase provides a generalized representation of possible configurations. However, it is still not ready for rendering with the Helm template (i.e., processing conditionals, loops, and placeholders in the template). The rendering process requires that only one value is indicated from the configuration space of enumerative fields in the schema.
To address this, \toolname{} performs the rendering multiple times, by exploring different combinations of values in enumerative fields. Each combination leads to a variant of the schema (\textit{values variant}). 

At each iteration, \toolname{} replaces enumerative placeholders in the schema with one of their valid values, while preserving placeholders for non-enumerative fields and constant fields.
To avoid combinatorial explosion, \toolname{} only explores a subset of configurations, such that each valid value for an enumerative field is covered in at least one generated variant. This process guarantees that the union of all variants covers all potential valid values from API requests, which should be allowed in the system by \toolname{}. More specifically, at each iteration $i$, the process generates a new values variant by replacing each enumerative field with its \textit{i}-th value. If an enumerative list has fewer options than the current iteration index, its last value is reused. The process iterates up to the length of the longest enumerative list. 
In the example of the schema in Figure \ref{fig:generated_values} (right), this process generates two values variants, one for each option in the \texttt{arch} enumerative field.

\vspace{2mm}
\noindent
\textbf{\textit{Rendering of manifests}}.
Once the \textit{values variants} are generated by the previous phase, they need to be translated into Kubernetes manifests. These manifests are concrete representations of resource specifications, by resolving conditionals and loops and using actual values. These manifests will be the basis for generating the final security policies. 

Each values variant is combined with the Helm template, using the \texttt{helm template} command. This command processes the template and values file to render a manifest. By the end of this phase, \toolname{} produces a set of Kubernetes manifests, capturing all permissible configurations for the resources required by the K8s operator.

\vspace{2mm}
\noindent
\textbf{\textit{Generation of validators}}.
The final step is to consolidate the generated manifests into a single \textit{validator}, a YAML schema that defines all allowable configurations for K8s resources. This validator supports the enforcement of fine-grained security policies, by serving as a reference for validating incoming API requests.

Manifests are grouped based on the resource type (\textit{kind}) (e.g., \textit{Pod}, \textit{Service}, \textit{Deployment}) to ensure the resulting validator is organized and easily navigable. Each group represents the allowed configurations for a specific Kubernetes resource type. Special placeholders (e.g., \texttt{string}, \texttt{int}, \texttt{hostIP}) from the manifests are retained to represent data types, enabling flexibility in configuration validation. Enumerative fields from multiple manifests are consolidated into arrays containing all valid values. Duplicate values are eliminated, while conflicting values are resolved by including all possible options in the array. Fields with constant, security-critical values (e.g., \texttt{securityContext.runAsNonRoot: true}) are directly carried over from the manifests without modification. This ensures that security best practices are enforced consistently across all configurations. Figure \ref{fig:generated_validator} shows a validator generated merging two manifests.

\Definition{
The proposed approach automates the generation of fine-grained security policies for specific K8s workloads, by generating a YAML policy validator that captures all their allowable configurations, based on the systematic analysis of their Helm charts. 
}

\begin{figure}[ht]
\begin{lstlisting}[language=yaml]
# Manifest sample 1             |
containers:                     |
- name: nginx                   |# Generated Validator
  image: nginx:latest           |containers:
  imagePullPolicy: IfNotPresent |- name: string
  ports:                        |  image: string
  - name: string                |  imagePullPolicy: 
    container:     IP           |  - IfNotPresent
# Manifest sample2              |  - Always
containers:                     |  ports:
- name: nginx                   |  - name: string
  image: nginx:latest           |    container: IP
  imagePullPolicy: Always       |
\end{lstlisting}
\caption{Policy Validator generated from two manifests.}
\label{fig:generated_validator}
\end{figure}

\begin{figure*}[t]
  \includegraphics[width=\textwidth]{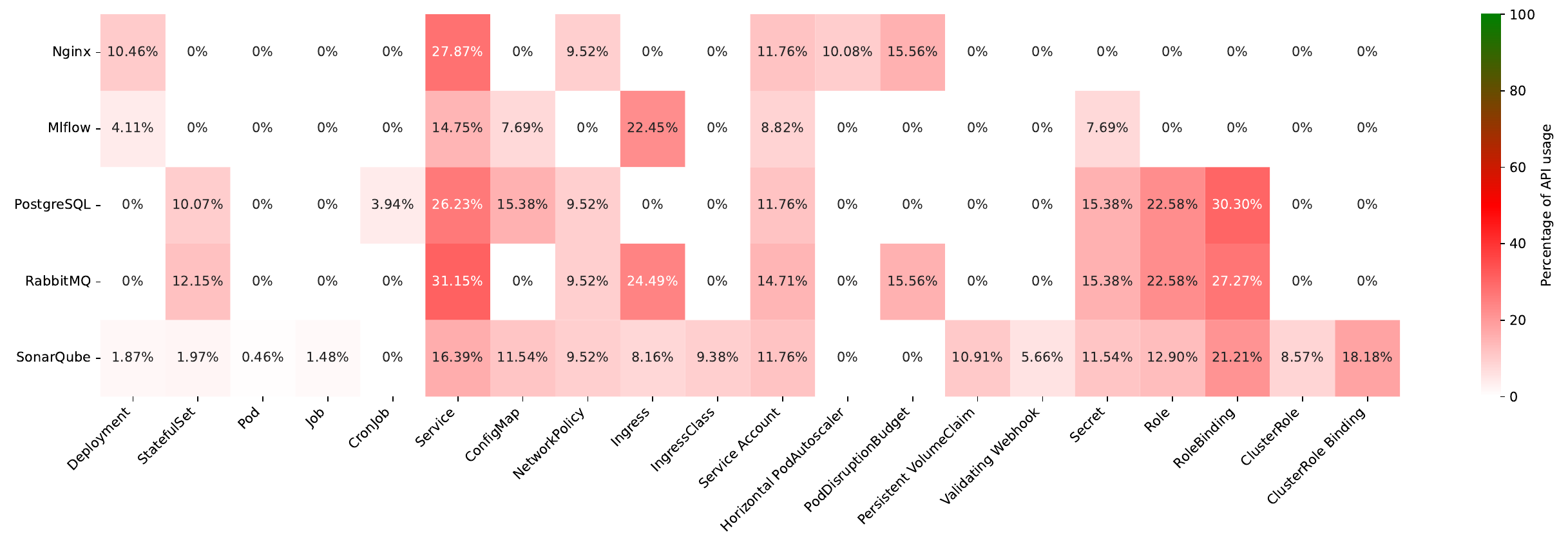}
  \caption{Percentage of API usage across workloads and endpoints.}
  \label{fig:API_usage}
\end{figure*}

\subsection{Enforcement of Security Policies}
To enforce security policies and apply the generated validators, \toolname{} employs Mitmproxy \cite{mitmproxy}. Mitmproxy is chosen for its capabilities in intercepting, inspecting, and modifying HTTP and HTTPS traffic, with support for SSL/TLS certificates. Mitmproxy is deployed as a Pod on each K8s control-plane node, positioned between the K8s API server and clients (e.g., \textit{kubectl, CI/CD pipelines, or Operators}). This ensures that all incoming API requests are intercepted and validated against the policy validator before reaching the API server.

In order to maintain a secure enforcement mechanism, the requests to the API server must not bypass the proxy (according to the \emph{Complete Mediation} design principle \cite{saltzer1975protection,smith2012contemporary}). To this aim, the API server is restricted to accepting only certificate-based trusted connections, allowing only the proxy with a valid certificate to connect. Furthermore, since client-to-API server connections are encrypted, clients must trust the CA certificate of the proxy to enable traffic interception and decryption for inspection. Thus, proper certificate management is crucial for secure operations.

The core validation mechanism is implemented as a Python-based Mitmproxy plugin, which loads the YAML policy validator. When Mitmproxy intercepts an HTTPS request, the plugin parses the request body to extract the Kubernetes object, including the resource type (\textit{kind}) and its specification fields, for validation. This validation process ensures that only authorized and correctly configured API requests are forwarded to the API server. The validation process operates as a hierarchical comparison, akin to a tree overlap, between the incoming manifest and the policy validator, interating across the requested K8s resources. 
The plugin extracts the \textit{kind} field from the request to identify the resource type and verifies its presence in the validator. 
Then, it ensures that only fields explicitly defined in the validator are present in the manifest and validates that each field's data type matches the expected type specified in the validator. 
Enumerative fields and security-sensitive fields are validated against their strict list of allowed values in the validator. 
If the request complies with the validator rules, it is forwarded to the K8s API server unchanged. 
Otherwise, the plugin blocks the request, returning an HTTP error response to the client. 
Violations are logged with details of the offending field and the reason for denial, enabling auditing and forensic analysis.


\section{Experimental Analysis}
This section evaluates \toolname{} across three dimensions. First, we quantify the attack surface exposed by the Kubernetes API and demonstrate how \toolname{} can reduce unnecessary exposure by restricting access to unused endpoints and fields. Second, using a catalog of misconfiguration-based attacks and CVE exploits targeting specific API fields, we assess the effectiveness of \toolname{} in mitigating these threats compared to native Kubernetes RBAC. Finally, we analyze the runtime overhead introduced by \toolname{} for API request validation.
 
\subsection{Experimental Setup}
\label{sec:setup}
We set up a K8s test environment replicating real-world deployment scenarios. The testbed includes a cluster using Kubernetes version 1.28.6, configured with a control-plane node and a worker node deployed on two distinct Ubuntu Linux 22.04.4 virtual machines. Both nodes are allocated $8$ vCPUs and $16$ GB of RAM, hosted on a machine with an \textit{Intel Xeon E5-1620} $3.70$ GHz CPU. \toolname{} is deployed on the control-plane node, besides the API Server, using a container with Mitmproxy version 10.2.2. 

This experimental analysis focuses on K8s Operators as a use case to demonstrate the feasibility and effectiveness of fine-grained workload-aware API enforcement. Since Operators are highly configurable, have extensive interactions with the K8s API \cite{henning2021reproducible}, and are typically deployed using Helm charts, they are a compatible target for validating \toolname{}. We select five Operators available on the Artifact Hub catalog \cite{artifact_hub}, representing diverse categories of workloads commonly deployed in K8s clusters, including databases (\textit{PostgreSQL} \cite{postgresql}), networking services (\textit{Nginx} \cite{nginx}), AI/ML applications (\textit{MLflow} \cite{mlflow}), data streaming (\textit{RabbitMQ} \cite{rabbitmq}), and security (\textit{SonarQube} \cite{sonarqube}).

\subsection{Quantifying K8s Attack Surface Exposure and Reduction}
The K8s API serves as the primary interface to a cluster, which the endpoints for users to query and modify resources. As a result, it represents a critical part of the attack surface for a cluster. We hypothesize that many workloads utilize only a small subset of this API, leaving a significant portion unnecessarily exposed and susceptible to exploitation. Reducing the accessibility of unused or unnecessary API endpoints and fields is a key objective of \toolname{}. In this experiment, we quantify the attack surface exposed by the Kubernetes API and evaluate how effectively \toolname{} can reduce it by limiting access to non-essential endpoints and fields. 

To quantify the attack surface, we conducted a static analysis of the K8s API. This process involved counting the total number of endpoints exposed by the K8s API Server, and cataloging the configurable fields available for each resource type. 
Next, we analyzed the selected Kubernetes Operators to understand how real workloads interact with the API endpoints. By examining the validators generated through \toolname{}, we identified the space of endpoints and fields that can potentially be used by each workload. This analysis provided a detailed understanding of K8s API utilization across different workloads and highlighted workload-specific behavior. The results are summarized in Figure \ref{fig:API_usage}, showing the percentage of fields utilized by each workload for each endpoint, relative to the total available fields. 

\begin{table}[b]
\caption{Attack Surface Reduction Achievable by KubeFence vs RBAC}
\label{tab:surface_reduction}
\centering
\begin{tabularx}{\columnwidth}{l*{4}{>{\centering\arraybackslash}X}}
\toprule
\textbf{Workload}    & \multicolumn{2}{c}{\textbf{Restrictable Fields}} & \multicolumn{2}{c}{\textbf{Attack Surface Reduction}} \\ 
\cmidrule(lr){2-3} \cmidrule(lr){4-5}
                     & \textbf{RBAC} & \textbf{KubeFence}   & \textbf{RBAC} & \textbf{KubeFence}        \\ \midrule
Nginx  & 3747 / 4882 & 4751 / 4882 & 76.75 \% & 97.32 \%   \\
Mlflow & 3883 / 4882 & 4826 / 4882 & 79.54 \% & 98.85 \%    \\
PostgreSQL  & 2906 / 4882 & 4711 / 4882 & 59.52 \% & 96.50 \%  \\
RabbitMQ & 3676 / 4882 & 4708 / 4882 & 75.30 \% & 96.44 \%    \\
SonarQube  & 1012 / 4882 & 4772 / 4882 & 20.73 \% & 97.75 \%  \\ 
\bottomrule
\end{tabularx}
\end{table}

Our findings revealed significant under-utilization of the K8s API by Operators, with numerous fields and endpoints remaining unused in practice. For instance, we observed that certain resources, such as \textit{Pod} and \textit{Job}, are entirely unused (i.e., 0\% usage), by a substantial number of workloads. 
Other resources, such as \textit{Service} and \textit{ServiceAccount}, are actively used by all workloads, even if many of their fields are left unused. It is still worth blocking these unused fields since they contribute to the attack surface, potentially serving as entry points for exploitation, despite offering no functional value to the workloads. However, these resources cannot be completely disabled, due to the frequent use of some of their fields. This leaves some residual risk of vulnerabilities in these APIs, as discussed in Section~\ref{sec:discussion}.

\begin{table*}[t]
  \caption{Catalog of K8s Malicious Specifications}
  \label{tab:misconfigurations_catalog}
  \centering
  \begin{tabular}{ p{0.02\linewidth} p{0.45\linewidth} p{0.4\linewidth} p{0.04\linewidth} } 
\hline
\textbf{ID} & \textbf{Exploit/Misconfiguration} & \textbf{Targeted API Field}  & \textbf{Ref.} \\ \hline

\rowcolor{white}
E1 & Activation of hostNetwork (\textit{CVE-2020-15257})   & hostNetwork &  \cite{CVE-2020-15257}\\

\rowcolor{gray!20}
E2 & Abusing \texttt{LoadBalancer} or \texttt{ExternalIPs} (\textit{CVE-2020-8554}) & externalIPs & \cite{CVE-2020-8554}\\

\rowcolor{white}
E3 & Command injection via \texttt{volume} (\textit{CVE-2023-3676}) & containers.volumeMounts.subPath & \cite{CVE-2023-3676} \\
\rowcolor{white}
& and \texttt{volumeMounts} & containers.volumes.subPath & \\

\rowcolor{gray!20}
E4 & Mount \texttt{subPath} on a file o \texttt{emptyDir} (\textit{CVE-2017-1002101}) & containers.volumeMounts.subPath & \cite{CVE-2017-1002101} \\ 

\rowcolor{white}
E5 & Absent Resource Limit (\textit{CVE-2019-11253}) & containers.resources.limits & \cite{CVE-2019-11253} \\

\rowcolor{gray!20}
E6 & Symlink exchange allow host filesystem access (\textit{CVE-2021-25741}) & container.command  & \cite{CVE-2021-25741} \\ 

\rowcolor{white}
E7 & Bypass of Seccomp Profile (\textit{CVE-2023-2431}) & containers.securityContext.seccompProfile.localhostProfile  & \cite{CVE-2023-2431} \\

\rowcolor{gray!20}
E8 & Privileged Containers (\textit{CVE-2021-21334}) & containers.securityContext.privileged & \cite{CVE-2021-21334}  \\ 

\rowcolor{white}
M1 & Activation of hostIPC & hostIPC & \cite{nsacisa2022hardening} \\ 

\rowcolor{gray!20}
M2 & Activation of hostPID & hostPID  & \cite{nsacisa2022hardening} \\ 

\rowcolor{white}
M3 & Use Readonly Filesystem & containers.securityContext.readOnlyRootFilesystem & \cite{nsacisa2022hardening} \\

\rowcolor{gray!20}
M4 & Running Containers as Root & containers.securityContext.runAsNonRoot & \cite{nsacisa2022hardening} \\
\rowcolor{gray!20}
& & containers.securityContext.runAsRootAllowed & \\

\rowcolor{white}
M5 & Allow Dangereous Capabilites to Containers & containers.securityContext.capabilities.add & \cite{nsacisa2022hardening} \\ 

\rowcolor{gray!20}
M6 & Escalated Privileges for Child Container Processes & containers.securityContext.allowPrivilegeEscalation & \cite{nsacisa2022hardening} \\ 

\rowcolor{white}
M7 & Custom SELinux \texttt{user} or \texttt{role} & containers.securityContext.seLinuxOptions.user  & \cite{nsacisa2022hardening} \\
 & & containers.securityContext.seLinuxOptions.role  & \\
   
   \bottomrule
  \end{tabular}
\end{table*}

To evaluate the attack surface reduction achievable by \toolname{} against RBAC, we analyzed the percentage of fields restrictable by each approach. RBAC restricts access to fields only when the entire resource type (API endpoint) is unused in the heatmap, meaning it lacks the granularity to filter individual fields within an allowed resource. In contrast, \toolname{} can enforce restrictions on any unused field, even within partially-used endpoints. This makes \toolname{} a strict superset of RBAC's enforcement, covering all fields RBAC could restrict while also providing additional reductions in the attack surface.

For the complete set of considered endpoints, we summed up the total configurable fields across all resources. For each workload, we computed the percentage of fields restrictable by RBAC and \toolname{} as a measure of the attack surface reduction potentially achievable by the two techniques. 
Table \ref{tab:surface_reduction} summarizes the results of this analysis. \toolname{} consistently achieves a higher reduction in attack surface across all workloads, with improvements of 20.57\%, 19.31\%, 36.98\%, 21.14\%, and 77.02\% across the five workloads, averaging 35\% compared to RBAC. These results highlight that RBAC achieves lower attack surface reduction for workloads requiring multiple endpoints, as it cannot blacklist partially-used resources. In contrast, \toolname{} can restrict unused fields within utilized endpoints, providing finer-grained protection.

\subsection{Catalog of Malicious Specifications}
\label{sec:misconfiguration_analysis}
As described in Section \ref{sec:motivation}, malicious API requests pose critical security risks, by exploiting specific fields to achieve privilege escalation, unauthorized access to critical resources, and misconfigurations that may led to degradation of cluster availability or reliability.

To evaluate \toolname{} against these threats, we built a catalog of 15 malicious specifications, comprising 7 misconfigurations and 8 malicious fields used by CVE exploits. These malicious specifications inject malicious values in Kubernetes manifests that can expose vulnerabilities or enable unsafe configurations, making them a practical subset for testing the effectiveness of \toolname{} in mitigating attacks. Table \ref{tab:misconfigurations_catalog} summarizes this catalog, identifying the targeted fields and providing references to their sources. 

This catalog was developed by analyzing prior research \cite{misconfiguration2023}, security blogs \cite{misc_1, misc_2, misc_3, misc_4}, CVE disclosure \cite{github_advisory}, and Kubernetes security guidelines \cite{sec_k8s_guidelines, nsacisa2022hardening}. Examples include enabling the \texttt{hostNetwork} field for host network sharing (CVE-2020-15257), exploiting \texttt{subPath} for host directory access (CVE-2017-1002101), and bypassing security profiles (CVE-2023-2431). We focus on CVEs from Section~\ref{sec:CVE_mapping} that are exposed to malicious specifications from the K8s API interface, as these align with our threat model and the scope of API-level enforcement. We exclude CVEs that are we are unable to reproduce due to strict environmental prerequisites, which fall outside our experimental setup. For instance, Kubernetes clusters are affected by CVE-2023-5528 only if they use an in-tree storage plugin for Windows nodes.

\subsection{Kubefence Effectiveness against RBAC}
Misconfigurations and CVE exploits pose significant security risks in K8s (see Section \ref{sec:motivation}). The native Kubernetes RBAC mechanism provides access control at the resource- and verb-level, but lacks the granularity to restrict individual fields within the resource specification. This experiment measures the effectiveness of \toolname{} compared to RBAC, by evaluating its ability to \textit{mitigate CVEs and misconfigurations}. To quantify this, we generate workload-specific policies for the five selected operators (listed in Section \ref{sec:setup}), and test whether \toolname{} can block API-based misconfigurations and CVE exploits. 

To test the enforcement mechanisms, we generated malicious API requests using our catalog of malicious specifications (Table \ref{tab:misconfigurations_catalog}). We inject the malicious fields in the catalog in resource types that support that malicious fields. For instance, the \texttt{spec.externalIPs} field is specific to \textit{Service} resources. Other fields apply to relevant K8s resources, such as to \textit{Pod} and higher-level abstractions like \textit{Deployment}, \textit{ReplicaSet}, \textit{StatefulSet}, and \textit{DaemonSet}. 

Legitimate resource configurations were retrieved from Operator manifests, and malicious fields were injected into this configuration to create 15 distinct malicious manifests for each operator. For example, Figure \ref{fig:malicious_manifest} shows how the misconfigured \texttt{runAsNonRoot} field is injected into a \textit{Deployment} resource from the Nginx Operator. 

\begin{figure}[t]
\begin{lstlisting}[language=yaml]
apiVersion: apps/v1
kind: Deployment
spec:
  template:
    spec:
      containers:
      - name: nginx
        image: testImage
        securityContext:
          runAsNonRoot: false
\end{lstlisting}
\caption{Example of a malicious YAML manifest}
\label{fig:malicious_manifest}
\end{figure}

These malicious manifests were then submitted to the K8s API while the respective workload-specific RBAC or \toolname{} policy was in place. This process simulates realistic attack scenarios where a malicious client attempts to exploit CVEs or misconfigurations through the K8s API interface. The effectiveness of each enforcement mechanism was measured by recording whether each CVE exploit or misconfiguration attempt was mitigated.

\begin{figure*}[t]
  \includegraphics[width=\textwidth]{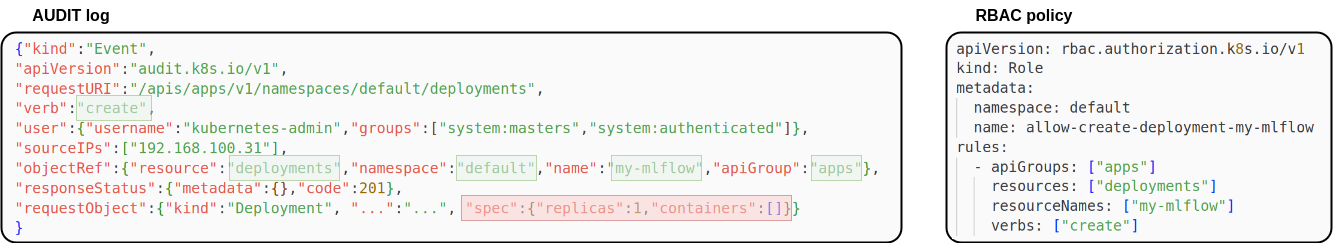}
  \caption{RBAC policy generated (on the right) from an audited \textit{create deployment} operation (on the left).}
  \label{fig:audit2rbac}
\end{figure*}

\textbf{\textit{Native K8s RBAC setup}.} We evaluated RBAC by configuring the K8s cluster with audit logging enabled, to capture API requests during the execution of an attack-free workload. Audit logs keep track of accesses to \textit{API endpoint}, the \textit{resource type} (e.g., Pods, Services), \textit{verb} (e.g., get, create, update, delete), and the resource specification in API requests. 
Then, the audit logs were processed with the \texttt{audit2rbac} tool \cite{audit2rbac}, which infers the minimum permissions required for a workload. This process generated five distinct YAML files, representing a tailored RBAC policy for each operator, based on the observed API interactions. 
Malicious manifests were applied to the cluster with RBAC policies in place. For each attack, we recorded the success or failure of the API request. 

\textbf{\textit{KubeFence setup}.} Then, we evaluated \toolname{} by generating fine-grained security policies tailored for each workload, by analyzing the configurations required by the workloads. These policies were enforced using our proxy placed between the clients and the K8s API Server.

The same malicious manifests used for testing RBAC were applied against the \toolname{} proxy. Each interaction of the Operators with the API server was intercepted by the proxy, which validated the API request against the workload-specific policy (i.e., the validator). 
Moreover, our logs report the denied actions, and the malicious fields that triggered by filtering.

\textbf{\textit{Experimental Results}}. Table \ref{tab:kubefence_comparison} reports the mitigated CVEs and misconfigurations by RBAC and \toolname{}, respectively. While RBAC did not block any of the attacks, \toolname{} successfully blocked all of them.

The results highlight that RBAC policies, even when tailored to workloads using tools like \textit{audit2rbac}, lack the granularity to enforce restrictions on individual fields within resource specifications. 
Figure \ref{fig:audit2rbac} illustrates an audit entry recorded during the deployment of the MLflow Operator, logging the creation of a \textit{Deployment} resource. The generated RBAC policy effectively defined access at the resource level, specifying resource \textit{kind}, \textit{namespace}, \textit{API group}, and allowed \textit{verbs}. However, it omitted critical parameter-level details, such as \texttt{spec} fields ``available'' in the audit logs.  This omission is not a limitation of \textit{audit2rbac}, but rather an inherent limitation of RBAC policies, which do not allow specification at this level of detail. As a result, RBAC failed to prevent attacks that exploit features not needed by the operators, such as enabling \textit{hostNetwork} or disabling \textit{runAsNonRoot}.

By contrast, \toolname{} successfully enforced fine-grained controls, blocking all attacks to misconfigurations and CVEs. For instance, it denied requests abusing the \texttt{subPath} field, as this parameter was not part of the configuration space defined in the Helm charts of the Operators. Furthermore, legitimate workload actions were unaffected, demonstrating the precision and reliability of \toolname{} in blocking unauthorized API request parameters without disrupting normal operations.

\begin{table}[t]
\caption{Mitigated CVEs and Misconfigurations by RBAC and KubeFence. }
\label{tab:kubefence_comparison}
\centering
\begin{tabularx}{\columnwidth}{l*{4}{>{\centering\arraybackslash}X}}
\toprule
\textbf{Workload}    & \multicolumn{2}{c}{\textbf{CVEs}} & \multicolumn{2}{c}{\textbf{Misconfigurations}} \\ 
\cmidrule(lr){2-3} \cmidrule(lr){4-5}
                     & \textbf{RBAC} & \textbf{KubeFence}   & \textbf{RBAC} & \textbf{KubeFence}        \\ \midrule
\textbf{PostgreSQL}  & \cellcolor{red!30}0 & \cellcolor{green!30}8 & \cellcolor{red!30}0 & \cellcolor{green!30}7 \\
\textbf{Nginx}       & \cellcolor{red!30}0 & \cellcolor{green!30}8 & \cellcolor{red!30}0 & \cellcolor{green!30}7 \\
\textbf{Mlflow}      & \cellcolor{red!30}0 & \cellcolor{green!30}8 & \cellcolor{red!30}0 & \cellcolor{green!30}7 \\
\textbf{RabbitMQ}    & \cellcolor{red!30}0 & \cellcolor{green!30}8 & \cellcolor{red!30}0 & \cellcolor{green!30}7 \\
\textbf{SonarQube}   & \cellcolor{red!30}0 & \cellcolor{green!30}8 & \cellcolor{red!30}0 & \cellcolor{green!30}7 \\ \bottomrule
\end{tabularx}
\end{table}

\subsection{KubeFence Overhead}
This section evaluates the runtime overhead introduced by \toolname{} compared to the native K8s RBAC. The focus is on the online phase, where API requests are inspected and forwarded to the API server, as this directly impact cluster operations. The offline phase of \toolname{}, which involves learning security policies, is excluded from this analysis as it does not affect runtime performance. 

We measured the round-trip-time (RTT) latency for processing the full set of API requests generated during the deployment of the five selected operators. These requests are issued by the \texttt{kubectl apply} command from the client, and include all interactions with the API server to configure the resources defined by the Operator. The RTT latency was defined as the total elapsed time from issuing the \textit{apply} command until the client received a response, indicating the API server had finished processing the requests. This experiment was conducted under two scenarios: first with native RBAC, and then with \toolname{}. All requests were benign, as this experiment focused on normal operations rather than attack scenarios. To ensure statistical significance, we repeated the process 10 times per workload and computed the average latency and standard deviation for both RBAC and \toolname{}. 
In addition, using the same workload, we evaluate the impact of \toolname{} on system resources. To this end, we measured the CPU and memory usage of the proxy container, reporting the average and standard deviation over 10 repetitions.

Table \ref{tab:latency_comparison} presents the average latencies and standard deviations for each workload, as well as the increase in latency introduced by \toolname{} over native RBAC.

\begin{table}[b] 
\caption{RBAC vs KubeFence Average Request Latency} 
\label{tab:latency_comparison} 
\centering 
\begin{tabularx}{\columnwidth}{Xlll}
\toprule
\textbf{Operators} & \textbf{RBAC RTT} & \textbf{\toolname{} RTT} & \textbf{Increase} \\ 
 & \textbf{(ms)} & \textbf{(ms)} & \textbf{(ms, \%)} \\ 
\midrule
MLflow      & $211.0 \pm 39.2$ & $237.6 \pm 37.5$ & $+26.6$ (12.61\%) \\ 
Nginx       & $168.4 \pm 25.7$ & $210.4 \pm 26.7$ & $+42.0$ (24.94\%) \\ 
PostgreSQL  & $178.1 \pm 16.1$ & $213.6 \pm 13.0$ & $+35.5$ (19.93\%) \\ 
RabbitMQ    & $242.9 \pm 16.6$ & $307.6 \pm 23.4$ & $+64.7$ (26.64\%) \\ 
SonarQube   & $385.9 \pm 14.0$ & $470.5 \pm 35.0$ & $+84.6$ (21.92\%) \\ 
\bottomrule 
\end{tabularx}
\end{table}

The results show that the additional latency introduced by \toolname{} remains minimal, with absolute increases ranging from $0.0266$ s to $0.0846$ s. Even in the worst case, where the overhead reaches $26$\%, the total RTT latency remains well below $0.5$ seconds, which is negligible for cluster management tasks such as workload deployment \cite{10533288, kubernetes_slo}. Furthermore, this overhead impacts only operations initiated by external actors interacting with the K8s control plane, such as managing deployments and querying resource modifications, while internal API interactions by K8s components remain unaffected by \toolname{}. Moreover, the application themselves (e.g., web resources served by Nginx) are unaffected. This minimal overhead is a worthwhile trade-off considering the enhanced attack mitigation capabilities and the attack surface reduction provided by \toolname{}. 

In addition, the impact of \toolname{} on system resources was minimal. CPU usage increased  only by $1.21\%$ ($\pm 0.04$), and memory consumption increased by $85.54$ MiB ($\pm 0.25$), which is a negligible overhead considering the security benefits.

\Definition{
       The experimental results demonstrate that \toolname{} effectively enhances Kubernetes security. It achieves significant attack surface reduction by restricting unused API fields and endpoints, addressing gaps in native RBAC that cannot enforce such fine-grained control. Moreover, \toolname{} successfully blocks all tested misconfigurations and CVE exploits by precisely validating API requests against workload-specific policies, ensuring robust protection against real-world threats. Despite introducing a small latency overhead during API request validation, the impact remains reasonable for operations initiated by external actors and does not affect internal K8s operations. These findings establish \toolname{} as a practical and efficient solution for securing K8s environments.
}

\section{Related Work}
\subsection{Static Analysis}
Kubernetes is widely adopted for its scalability and automation capabilities. However, its extensive configurability introduces significant risks of misconfiguration. Static analysis tools and methodologies have been developed to address these issues, focusing on pre-deployment detection in Kubernetes manifests and Helm Charts. 

Empirical studies \cite{shamim2021mitigating, bose2021under, misconfiguration2023} highlight the prevalence of Kubernetes misconfiguration and their security implications. Research analyzing Helm charts and Operators \cite{zerouali2023helm, xu2024empirical} identifies insecure defaults and outdated dependencies that elevate security risks. Broader efforts \cite{shamim2020xi, nsacisa2022hardening} emphasize adherence to best practices, including RBAC or network segmentation, but these approaches rely on manual intervention and lack runtime adaptability. 

Static analysis tools, such as KubeLinter \cite{kube_linter}, Polaris \cite{polaris}, Checkov \cite{checkov}, KICS \cite{kiks}, and SLI-Kube \cite{misconfiguration2023} identify misconfigurations using predefined rules. Graph-based methods such as KGSecConfig \cite{haque2022kgsecconfig} and \cite{blaise2022stay, dellimmagine2023kubehound} automate secure cluster configuration or provide security assessment of configurations. Generative approaches like GenKubeSec \cite{malul2024genkubesec} and \cite{minna2024analyzing, lanciano2023analyzing} leverage large language models to identify and remediate misconfigurations. Despite their utility, these tools operate pre-deployment, leaving systems exposed to runtime threats, such as malicious API requests targeting unused or overly permissive fields. 

RBAC misconfiguration detection tools, such as EPScan \cite{gu2024epscan}, focus on identifying excessive permissions but lack mechanisms to enforce fine-grained control at runtime. These approaches are inadequate for addressing threats where unused API fields can be exploited.

In contrast, \toolname{} bridges this gap by dynamically enforcing workload-specific API policies at runtime, reducing the attack surface through fine-grained control of resource specifications.

\subsection{Container Security}
Container-centric security solutions primarily address runtime behaviors of containers, targeting process execution and interactions within the host environment. 

Runtime monitoring tools, such as system call monitoring \cite{kitahara2020highly}, detect anomalous container behavior but require extensive pre-configuration and rely on heuristic models. ProSPEC \cite{kermabon2022prospec} predicts potential breaches through proactive policy enforcement. Tools like Kub-Sec \cite{zhu2022kub} and KubeArmor \cite{kube_armor} generate security profiles for pods, enforcing the principle of least privilege. Sysdig Falco \cite{falco} and similar tools monitor container operations based on predefined rules. Tools like eBPF \cite{eBPF} and Seccomp \cite{seccomp} can be used to whitelist allowable syscalls from containers. However, these approaches focus on behaviors occurring within workload containers, rather than K8s API requests. In principle, syscall whitelisting can detect some of the malicious behaviors that may occur after exploiting K8s vulnerabilities, e.g., executing privileged system calls. However, it is quite challenging to define policies for finer-grain filters on system calls, since it would need complex static/dynamic analysis of programs, or manual definitions. Moreover, some malicious behaviors can still escape system call policies, such as by abusing privileges that are available to the application.

\toolname{} shifts the focus to API-level security, complementing container security solutions. By dynamically generating fine-grained policies tailored to workload configurations, it secures the Kubernetes control plane against threats arising from unused API fields and misconfigurations, offering comprehensive protection for the Kubernetes ecosystem.

\subsection{REST API Security}
Securing REST APIs have been a key focus in web and application security, with numerous methodologies proposed for generating and enforcing access control policies to mitigate unauthorized access and enhance runtime security. Languages and frameworks like RestPL \cite{luo2016restpl} facilitate precise and flexible policy definitions for RESTful APIs, emphasizing request-level granularity but remaining confined to static pre-deployment configurations. Similarly, Jayathilaka et al. \cite{jayathilakarest} propose a framework for enforcing API security policies in cloud platforms, which ensures backward compatibility and enforces best practices during API deployment. Atlidakis et al. \cite{atlidakis2020checking} extend REST API security through property checking, fuzzing, and runtime monitoring, while a more recent framework by Khan et al. \cite{khan2024framework} focuses on detecting and mitigating vulnerabilities in REST APIs by integrating reverse proxy techniques to identify and prevent attacks like SQL injection and XSS in real-time. 

While effective for traditional REST APIs, these solutions lack mechanisms for dynamically adapting policies on the rich configurations and nested specifications unique to Kubernetes APIs. In contrast, \toolname{} extends these principles by generating fine-grained security policies tailored to Kubernetes workloads.

\section{Discussion}
\label{sec:discussion}

\vspace{2mm}
\noindent
\textbf{\textit{Extensibility beyond Helm}}. 
The current implementation of \toolname{} focuses on Helm-based workloads, using Helm templates to generate API security policies. While effective, this limitations its applicability to Helm deployments. However, the methodology of analyzing manifests to derive workload-specific security policies can be easily extended to other deployment mechanisms, such as Kustomize or raw YAML manifests. By adapting the parsing and policy generation processes, \toolname{} can ensure consistent security enforcement across diverse deployment workflows. 

\vspace{2mm}
\noindent
\textbf{\textit{Scope of attack surface hardening}}. 
The primary objective of \toolname{} is to \textit{reduce the Kubernetes attack surface}, denying unnecessary and risky features on a per-workload basis. Despite this, it does not claim to eliminate CVE exploitability or misconfiguration risks entirely. The solution leverages the client configuration space to infer which API endpoints and fields allow, and best practices guidelines to infer which critical field to lock to safe values. This significantly mitigates opportunities for attackers to exploit malicious configurations irrelevant to the workload. 

\toolname{} is based on the idea of blocking code not used by common workloads, which can be difficult to apply for some users. 
It is possible that uncommon, but legitimate workloads are blocked by such restrictive security policy. In general, false positives are a challenge for any filtering approach, as the same issue is faced by admins that manage firewalls, IDS, and similar tools. \toolname{} mitigates false positives by tailoring policies to K8s operators, which are becoming a popular approach to manage clusters. In such use cases, features can be restricted with very high accuracy. However, in other use cases, it may be difficult to anticipate which features should be allowed. Still, some enterprises may still want to block uncommon workloads, and enable them only after more careful scrutiny.

It is also possible that \toolname{} does not restrict interfaces that are prone to vulnerabilities, in the case that these interfaces are used by legitimate workloads, in order not to disrupt them. These interfaces represent a residual risk, which has to be handled through other complementary strategies. 
One approach is to adopt anomaly detection methods on API calls, which can identify misuses and exploitation attempts of the features \cite{ARMO_kubescape}. Another strategy is to perform more thorough testing, such as fuzzing \cite{dommaraju2024erroneous}, to identify vulnerabilities in the residual attack surface.

\toolname{} does not validate the functional correctness of Helm charts when they introduce unnecessary features or omit required ones. Instead, it enforces the stated resource definitions as provided. Ensuring correctness in such cases is an orthogonal problem and fall outside the scope of \toolname{}. External YAML validation tools (e.g., KubeLinter \cite{kube_linter}, Checkov \cite{checkov}) can be used before policy generation to address this problem.

Furthermore, \toolname{} does not address risks arising from compromises in the Kubernetes Operator catalog through supply chain attacks, where malicious Operators could inject unsafe configurations. In such cases, the responsibility for addressing these risks lies with workload developers.

\vspace{2mm}
\noindent
\textbf{\textit{Performance Optimizations}}. 
While the overhead introduced by \toolname{} is negligible for most cluster management tasks, with latency increases ranging from $0.0266$ to $0.0846$ s, it could still impact performance-critical of real-time deployment scenarios. The current implementation relies on a proxy Pod to intercept, validate, and forward external API requests to the API server, which adds network latencies. To address this, \toolname{} could be integrated directly into the API server, eliminating the forwarding delays and leaving only the validation cost. Although it requires modifications to the API server codebase, this approach offers a viable path to optimize enforcement efficiency for more demanding use cases.

\section{Conclusion}
The extensive Kubernetes API and reliance on coarse-grained RBAC leave clusters vulnerable to misconfigurations and CVE exploits. This paper introduced \toolname{}, a proxy-based enforcement mechanism that enhances Kubernetes security by automatically generating and enforcing fine-grained API security policies tailored to workloads, effectively reducing the attack surface and mitigating insider threats. 

\section*{Acknowledgment}
We are grateful to our shepherd François Taïani and to the anonymous reviewers for their feedback. This work was supported by the projects GENIO (CUP B69J23005770005) funded by MIMIT, and ``IDA—Information Disorder Awareness'' funded by the European Union-Next Generation EU within the SERICS Program through the MUR National Recovery and Resilience Plan under Grant PE00000014.

\bibliographystyle{IEEEtran}


\end{document}